\documentclass[10pt,journal]{IEEEtran}
\usepackage{dsfont}
\usepackage{amssymb}
\usepackage{amsmath}
\usepackage{cite}
\usepackage{graphicx}
\usepackage{booktabs}
\usepackage{nccmath}

\author{\IEEEauthorblockN{Yun Wei$^{*1}$, Chuanyi Ji$^1$, Floyd Galvan$^2$, Stephen Couvillon$^2$, George Orellana$^2$, James Momoh$^3$} \\
\IEEEauthorblockA{$^1$
Georgia Institute of Technology, Atlanta, GA 30332--0250 \\
$^2$Entergy Services, Inc., New Orleans, LA 70053 \\
$^3$
Howard University, NW Washington, DC, 20059  \\
Email: \{ywei30, jichuanyi\}@gatech.edu}}
\title{Non-Stationary Random Process for Large-Scale Failure and Recovery of Power Distributions}
\begin{document}
\maketitle

\begin{abstract}

A key objective of the smart grid is to improve reliability of utility services to end users. This requires strengthening resilience of distribution networks that lie at the edge of the grid. However, distribution networks are exposed to external disturbances such as hurricanes and snow storms where electricity service to customers is disrupted repeatedly. External disturbances cause large-scale power failures that are neither well-understood, nor formulated rigorously, nor studied systematically. This work studies resilience of power distribution networks to large-scale disturbances in three aspects. First, a non-stationary random process is derived to characterize an entire life cycle of large-scale failure and recovery. Second, resilience is defined based on the non-stationary random process. Close form analytical expressions are derived under specific large-scale failure scenarios. Third, the non-stationary model and the resilience metric are applied to a real life example of large-scale disruptions due to Hurricane Ike. Real data on large-scale failures from an operational network is used to learn time-varying model parameters and resilience metrics.

\end{abstract}


\section{Introduction}\label{sec:Intro}

Our power grid is a vast interconnected network that delivers electricity from suppliers to consumers. At the edge of the grid lies power distribution networks \cite{Kaplan09}. Power distribution networks provide medium or low voltages to residence and organizations, and thus serve a unique role of connecting the grid to end users.

Distribution networks consist of "leaf nodes" of the energy infrastructure and are thus susceptible to external disturbances. For example, natural disasters repeatedly cause devastating destructions and service disruptions to distributions networks \cite{Doereport}\cite{Albert04}. There were 6 major hurricanes and more than 10 snow and ice storms occurred in north America in the past 5 years \cite{Wiki}. Each natural disaster disrupted power services to more than 500,000 customers for days \cite{Wiki}. Large-scale power outages are more prevalent and damaging in developing countries, whose energy demands are rapidly increasing but power infrastructures are still in development \cite{Rudnick11}.

A fundamental research problem pertaining to the real problem is the resilience of power distribution to large-scale external disruptions. Resilience corresponds to the ability of distribution networks to withstand external disturbances and to recover rapidly from failures. Up to date, tremendous attention has been directed to resilience of the core that consists of major power generation and transmission systems of high voltages \cite{Pagani11}\cite{Hines09}\cite{Scada04}. As an external disturbance often affects power distribution networks in a wide geo-graphical span, the service disruptions usually remain local at the edge \cite{DOE}. The resilience of the edge, however, is understudied \cite{Kuh-pc}. As the demand on energy is growing, the edge of the grid is becoming thicker. For example, a utility provider often serves millions of customers in America. Damages from external disturbances on power distribution can thus result in a profound impact to a large number of users. Hence, the issue of resilience of power distribution networks is much needed to be investigated.

Empirical studies have been conducted on how to access damages from large-scale power outages (see \cite{Doereport1} and reference therein). Monitoring systems have been developed and used in power industry to respond to failures due to natural disasters (see \cite{Entergy11} as an example). These empirical methods predict the degree of damages, e.g., the maximum number of outages upon a natural disaster \cite{Doereport}, often through observations and experiences. Quantifiable approaches are lacking and needed to leverage practical experiences to general principles on resilience of power distribution \cite{Rudnick11}\cite{Hooke07}\cite{Kuh-pc}. In other words, resilience, as a concept about an overall power distribution network, needs to be learned systematically from real data.

Two research challenges emerge. One is what to learn from real data. Real data consists of samples on responses of a distribution network to external disturbances. External disturbances such as hurricanes often occur suddenly and unpredictably. The resulting large-scale failure and recovery at distribution networks thus exhibit random and dynamic behavior. For example, recovery depends on multiple random factors such as environmental conditions, available resources and preparation. Prior work models network failures and recoveries using finite state Markov processes \cite{Liu04}. These models belong to the general birth-and-death process \cite{Ross10} which include randomness but assume stationary failure and recovery. Non-stationary failure such as a non-homogeneous Poisson point process is provided in the context of $Mt/G/\infty$ queues \cite{Massey93Poiss}\cite{Massey93} with time varying Poisson parameters. General time-dependent infinite-server arrival/service processes are studied in \cite{Nelson04a}\cite{Nelson04b}. However, non-stationary recovery is rarely included in network resilience. In fact, it is an open issue how to characterize large-scale non-stationary failure and non-stationary recovery for power distribution networks.


The second challenge is how to define and characterize resilience for an entire non-stationary life-cycle of failure and recovery. Prior work provides general discussions that resilience should include multiple attributes, such as both failure and recovery \cite{AlKuwaiti06}\cite{Ellison97}\cite{Lin06}. However, most prior works mainly study resilience in terms of maintaining services, and thus consider failures only \cite{Liew94}. In fact, recovery is rarely included in defining resilience of power distribution networks. It is an open issue how to derive resilience metrics for characterizing non-stationarity in both failure and recovery.

Hence it is necessary to formulate, from ground up, an entire life cycle of large-scale failure and recovery. This can prevent us from choosing a model and/or a learning approach subjectively. We first develop a problem formulation of large-scale failures at the finest level of network nodes based on temporal-spatial stochastic processes. However, as an individual external disturbance results in one "snapshot" of network responses, information (data) from one snapshot is insufficient to completely specify such temporal-spatial model \cite{Kant}. We thus derive temporal models of an entire distribution network by aggregating spatial variables. The resulting temporal process models an entire non-stationary life-cycle of large-scale failures. The model applies to general failure process that can be dependent and with an arbitrary distribution. Two distinct recovery-characteristics emerge from our model. One is infant recovery that reflects the ability to recover rapidly from failures. The other is aging recovery that corresponds to prolonged failures. We define the resilience as the probability of infant recovery for a power distribution network. We then derive analytical expressions for special cases of failure and recovery.

The model and the resilience metric are studied in a real life example of large scale service disruptions of power distribution. Power failures occurred during a major natural disaster, Hurricane Ike in 2008. Pertinent resilience parameters are learned using real data from an operational distribution network.

The rest of the paper is organized as follows. Section \ref{sec:Background} provides background knowledge and an example of large-scale failures at power distribution networks. Section \ref{sec:StochasticModel} develops a problem formulation of failure-and-recovery processes. Section \ref{sec:FRProc} characterizes non-stationary failure and recovery individually and jointly; then derives analytical expressions for special cases of failure and recovery. Section \ref{sec:Resilience} defines network resilience based on the non-stationary model. Section \ref{sec:RealData} learns pertinent resilience parameters using large-scale real data. Section \ref{sec:Conclusion} discusses our findings and concludes the paper.

\section{Background and Example}\label{sec:Background}

A power distribution network is at the edge of the grid from substations to users. A power distribution network consists of components including substations, feeders, transformers, poles, and transmission lines, and meters.

A large number of such devices in a distribution network are often in the open, and thus susceptible to natural disasters such as hurricanes, ice and snow storms. For example, a fallen pole can cause a short circuit, and other devices to fail subsequently. An external disturbance such as a hurricane can cause a large number of failures. The failures interrupt electricity service to end users. A large-scale external disruption can affect one or many distribution networks in a wide geo-graphical area. Failure recovery is often done by dispatching crews to the field. Promptness of failure recovery thus depends on environmental constraints, preparedness, and resources.

\begin{figure}
\begin{center}
\includegraphics[width=0.5\textwidth]{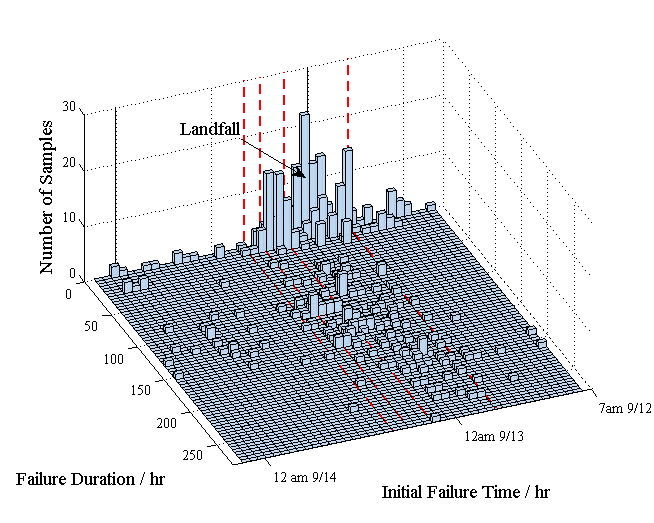}
\caption{Empirical distribution of failure durations in 3D.}
\label{fig:HistPDur}
\end{center}
\end{figure}

To gain intuition on an entire life cycle of failure and recovery, we consider an example of large-scale power failures occurred during Hurricane Ike. Hurricane Ike had a landfall in 2008 and affected densely populated areas in Texas and Louisiana. Figure \ref{fig:HistPDur} shows a histogram on failure occurrence time and duration at an operational distribution network before, during and after the hurricane. The example provides the following observations:

(a) Both failure occurrence and recovery are time-varying, i.e., non-stationary.

(b) Samples on failure occurrence time and duration are not identically distributed. Instead, recovery time characterized by failure duration is different for failures occurred at different time.

This shows that failure occurrence and recovery are statistically dependent, and non-stationary.



\section{Stochastic Model}\label{sec:StochasticModel}

We formulate large-scale failure and recovery based on non-stationary random processes. We begin with the detailed information on nodal statuses in a distribution network. We then aggregate the spatial variables of nodes to obtain temporal evolution of failure and recovery of an entire network.

\subsection{Failure and Recovery Probability}

A temporal-spatial random process provides a theoretical basis for modeling large-scale failures at the finest scale of nodes. The temporal variable is time $t$. The spatial variable is the location of a node in the grid. Here, nodes can be any components in a distribution network such as substations, feeders, hubs, transformers, transmission lines, and power circuits. A shorthand notation $i$ is used to specify both the index of node $i$ and its corresponding location, where $i\in S=\{1,2,...,n\}$ for a power distribution network with $n$ nodes.

Let $X_i(t)$ be the status of the $i$-th node at time $t>0$ for $1 \le i \le n$. $X_i(t)=1$ if the $i$-th node is in a failure mode. $X_i(t)=0$ if the node is in normal operation. For a distribution network, an example of a node failure includes a failed circuit, a fallen pole, a broken link, and an non-operational substation.

Failures caused by external disturbances exhibit randomness. Whether and when a node fails is random. Whether and when a failed node recovers is also random. Hence, random processes can be used to characterize failure and recovery for all nodes in a network.

Given time $t>0$, $P\{X_i(t+\tau)=1\}$ characterizes the probability that node $i$ is failed in the near future $t+\tau$, where $\tau>0$ is a small time increment. Assume a node changes state, i.e., from failure to normal and vice versa. Then for the $i$th node, $1 \le i \le n$,
\begin{equation}
\begin{split}
&P\{X_i(t+\tau)=1\}-P\{X_i(t)=1\} \\
=&P\{X_i(t+\tau)=1,X_i(t)=0\} \\
&-P\{X_i(t+\tau)=0,X_i(t)=1\}. \\
\end{split}
\label{eq:StateTrans1}
\end{equation}

Eq.\ref{eq:StateTrans1} provides a model for an individual node in a network. The model includes Markov temporal dependence and spatial dependence among nodal statuses. Such a model can be applied to a heterogeneous grid where nodes experience in general different failure and recovery processes. There are $n$ such temporal-spatial equations for $n$ nodes in a distribution network. The $n$ equations together form a temporal-spatial model for a network.

\subsection{Temporal Process}

When large-scale outages are caused by individual external disturbances, information available is from "snapshots" of temporal spatial network statuses. A snapshot corresponds to spatial-temporal nodal statuses with respect to one external disturbance, e.g., one hurricane. As there are usually only a few such snapshots available, information is insufficient for specifying a complete temporal spatial model at the node level. However, spatial variables can be aggregated out from Eq.\ref{eq:StateTrans1}, making the information sufficient for temporal characteristics, where

\begin{equation}
\begin{split}
&\sum_{i \in S} P\{X_i(t+\tau)=1\}-\sum_{i \in S} P\{X_i(t)=1\} \\
=&\sum_{i \in S} P\{X_i(t+\tau)=1,X_i(t)=0\} \\
&-\sum_{i \in S} P\{X_i(t+\tau)=0,X_i(t)=1\}. \\
\end{split}
\label{eq:StateTrans2}
\end{equation}

Furthermore, the probability can be related to an indicator function,
\begin{equation}
P\{X_i(t)=1\}=E\{I[X_i(t)=1]\}.
\label{eq:ind}
\end{equation}
Then we can define a temporal process as follows.

\indent

\textbf{Definition}: A temporal processes $\{N(t) \in \mathds{N},t>0\}$ is a special case of the temporal-spatial process where the spatial variables ($i$'s) are aggregated for all nodes in a network. $N(t)$ is the number of nodes in failure state at time $t$,
\begin{equation}
N(t)=\sum_{i \in S} I[X_i(t)=1],
\label{eq:frpDef}
\end{equation}
where $I(A)$ is an indicator function, i.e., $I(A)=1$ if event A occurs, and $I(A)=0$ otherwise.

\indent

Combining Equations \ref{eq:StateTrans2}, \ref{eq:ind}, and \ref{eq:frpDef}, we have,
\begin{equation}
\medmath{E\{\Delta N(t)\}=\sum_{i \in S} P\{X_i(t+\tau)=1\} - \sum_{i \in S} P\{X_i(t)=1\}},
\label{eq:frpExp}
\end{equation}
where $\Delta N(t)=N(t+\tau)-N(t)$. Hence, an expected increment of the number of failed nodes in a network equals to the total change of the aggregated probabilities on nodal statuses.

An increment $\Delta N(t)$ in the total number of nodes in failure state can result from either newly failed or newly recovered nodes. To further characterize the temporal process $N(t)$, we define a failure process and a recovery process respectively.

\indent

\textbf{Definition}: Failure and recovery processes: Failure process $\{N_f(t) \in \mathds{N},t\ge 0\}$ is the number of failures occurred up to time $t$. Recovery process $\{N_r(t) \in \mathds{N},t\ge 0\}$ is the number of recoveries occurred up to time $t$.

\indent

Assume $\tau>0$ is sufficiently small so that at most one failure occurs during $(t, t+\tau)$, 
then an increment on the number of failures satisfies
\begin{equation}
E\{\Delta N_f(t)\}= \sum_{i \in S} P\{X_i(t+\tau)=1,X_i(t)=0\} ,
\label{eq:fpExp}
\end{equation}
where $\Delta N_f(t)=N_f(t+\tau)\} - N_f(t)$. Similarly, for a sufficiently small $\tau$, it can be assumed that at most one recovery occurs during $(t,t+\tau)$. Then,
\begin{equation}
E\{\Delta N_r(t)\}= \sum_{i \in S} P\{X_i(t+\tau)=0,X_i(t)=1\} ,
\label{eq:rpExp}
\end{equation}
where $\Delta N_r(t)=N_r(t+\tau)\} - N_r(t)$. Hence, Eq.\ref{eq:StateTrans2} is simplified as,
\begin{equation}
E\{\Delta N(t)\} = E\{\Delta N_f(t)\}- E\{\Delta N_r(t)\}.
\label{eq:StateTrans3}
\end{equation}
Furthermore, we assume at time $t_0=0$, $N(t)=0$, $N_f(t)=0$, and $N_r(t)=0$. Aggregating increments in Eq.\ref{eq:StateTrans3} from $0$ to $t$, we have,
\begin{equation}
E\{N(t)\} = E\{N_f(t)\}- E\{N_r(t)\}.
\label{eq:FRE}
\end{equation}

Hence, the expected number of nodes in failure state equals to the difference between the expected failures and the expected recoveries. This characterizes how the status of a distribution network changes from the present to the near future.

For practicality, the empirical processes as the sample means $\hat N(t)$, $\hat N_f(t)$, and $\hat N_r(t)$ can be used to estimate the true expectations $E\{N(t)\}$, $E\{N_f(t)\}$, and $E\{N_r(t)\}$, respectively. Eq.\ref{eq:FRE} can then be represented by the empirical processes,
\begin{equation}
\hat N(t) = \hat N_f(t) - \hat N_r(t).
\label{eq:empFRE}
\end{equation}

The empirical processes and the failure-and-recovery equation allow learning from field data, which shall be elaborated in Section \ref{sec:RealData}.

\section{Non-Stationary Failure and Recovery}\label{sec:FRProc}

We now focus on the temporal processes to derive non-stationary characteristics on failure and recovery individually and jointly. We derive close form expressions for special cases of failure and recovery processes. Our study reveals pertinent parameters which shall be used to quantify resilience in Section \ref{sec:Resilience}.

\subsection{Failure Process}

Let $\lambda_f(t)$ be the intensity function of the failure process. $\lambda_f(t)$ is the expected number of new failures per unit time at epoch $t$, i.e.,
\begin{equation}
\lambda_f(t)=\lim_{\tau \rightarrow 0} {{ E\{N_f(t+\tau)-N_f(t)\}} \over {\tau}}.
\label{eq:FRateDef}
\end{equation}
$\lambda_f(t)$ is also referred to as the rate function of the failure process $N_f(t)$. The larger $\lambda_f(t)$ is, the more failures occur in a unit time duration. Hence, failure rate quantifies the intensity of failure occurrence. An non-stationary failure process has a time-varying intensity function $\lambda_f(t)$. Assuming a failure process begins at $t=0$, we have
\begin{equation}
E\{N_f(t)\}=\int_0^t \lambda_f(v)dv.
\label{eq:fp}
\end{equation}

The probability density function $f(t)$ of failure occurrence time at $t$ can be obtained from $\lambda_f(t)$, where
\begin{equation}
\begin{split}
f(t)=& \lim_{\epsilon \rightarrow 0} \frac{E\{N_f(t+\epsilon)-N_f(t)\}}{\lim_{v \rightarrow \infty} N_f(v)} \frac{1}{\epsilon} \\
    =& \frac{\lambda_f(t)}{ \int_0^{\infty} \lambda_f(s)ds }, \\
\end{split}
\label{eq:pdfFP}
\end{equation}

As a special case of a general failure process, $\lambda_f(t)$, as the first moment, can completely determines the failure process $N_f(t)$. Consider an independent failure process where the number of failures $N_f(t)$ is a counting process with independent increment $\Delta N_f(t)$. Among many such random processes, non-homogeneous Poisson process (NHPP) \cite{Ross10} captures the non-stationary nature of failures in a parametric form. The parameter is the intensity function, $\lambda_f(t)$. The non-stationarity refers to a common characteristic of large-scale external disruptions where power failures occur at different intensity at different time. The definition of a NHPP, applied to a failure process, is provided below.

\indent

\textbf{Definition} \cite{Ross10}: A continuous time counting process $\{N_f(t),t\ge 0\}$ is a Non-Homogeneous Poisson Process with a time-varying rate function $\lambda_f(t),t\ge 0$, if $N_f(t)$ satisfies the following conditions:
\begin{itemize}
\item $N_f(0)=0$;
\item $\{N_f(t),t>0\}$ has independent increments;
\item $P\{N_f(t+\tau)-N_f(t)\ge 2\}=o(\tau)$;
\item $P\{N_f(t+\tau)-N_f(t)=1\}=\lambda_f(t)\tau+o(\tau)$.
\end{itemize}

\indent

In Section \ref{sec:RealData}, we shall show using real data that a failure process $\{N_f(t)\}$ from a hurricane follows a NHPP with $\lambda_f(t)$ as the failure rate function.

\subsection{Recovery Process}

We now define intensity function $\lambda_r(t)$ for a recovery process. $\lambda_r(t)$ is the expected number of new recoveries per unit time at epoch $t$,
\begin{equation}
\lambda_r(t)=\lim_{\tau \rightarrow 0} {{ E\{N_r(t+\tau)\} - E\{N_r(t)\}} \over {\tau}}.
\label{eq:RRateDef}
\end{equation}
$\lambda_r(t)$ is also referred to as the rate function of the recovery process $N_r(t)$.

An non-stationary recovery process has a time-varying intensity function, i.e., $\lambda_r(t)$. Assuming the temporal failure process begins at $t_0=0$, we have
\begin{equation}
E\{N_r(t)\}=\int_0^t \lambda_r(v)dv.
\label{eq:fp}
\end{equation}

The recovery rate characterizes how rapidly recovery occurs, which is measured by failure duration $D$. For an non-stationary recovery process, a failure duration depends on when a failure occurs as illustrated in Figure \ref{fig:HistPDur}. Such non-stationarity of recovery is characterized by $g(d|t)$ which is a conditional probability density function of failure duration $D$ given failure time $T$. For a given threshold $d_0>0$, the conditional probability that a duration is bounded by $d_0$ for failures occurred around time $t$ is

\begin{equation}
P\{D < d_0|t\}=\int_0^{d_0} g(v|t) dv,
\label{eq:pd}
\end{equation}
where $D$ is a random failure duration.

When $d_0$ is sufficiently small, this probability characterizes rapid recovery that occurs shortly after failures. For a given $d_0$, the larger $P\{D < d_0|t\}$ is, the more dominating the rapid recovery is. Given desired value of probability $P\{D < d_0|t\}$, the smaller $d_0$ is, the more dominating the rapid recovery is.

Rapid recovery is referred to as infant recovery. This terminology is borrowed from infant mortality in survivability analysis \cite{Hosmer08}. Infant recovery is a desirable characteristic of the smart grid. In contrast, slow recovery is referred to as aging recovery in analogous to aging mortality \cite{Kalbfleisch02}. Infant and aging recovery can be formally defined as follows.

\indent

\textbf{Definition}: Infant and aging recovery: Let $d_0>0$ be a threshold value. If a node remains in failure for a duration less than $d_0$; a recovery is an infant recovery. Otherwise, the recovery is aging recovery. Infant recovery is characterized by $P\{D < d_0|t\}$. Aging recovery is characterized by $P\{D > d_0|t\}$.

\indent

Note here $P\{D < d_0|t\}$ is a function of failure occurrence time. As we shall show through a real-life example in Section \ref{sec:RealData}, failures occurred at different time may experience infant and/or aging recovery of different degrees, showing the non-stationarity of a recovery process.

\subsection{Joint Failure-and-Recovery Process}

A joint failure and recovery process characterizes an entire life cycle of a failure-and-recovery process (FRP), and represents the total number of nodes $N(t)$ in failure state at time $t$ (Eq.\ref{eq:frpDef}). The expected number of nodes in failure can be written in terms of rate functions,
\begin{equation}
E\{N(t)\}=\int_{0}^t [\lambda_f(v)-\lambda_r(v)]dv.
\label{eq:frp}
\end{equation}

Failure-and-recovery process can be viewed as an alternative form of the birth-and-death process \cite{Ross10}. However, commonly-used birth-and-death processes have a stationary distribution of failure duration and assume independence between failure occurrence $t$ and failure duration $d$. Here, these two assumptions are not needed. This implies that failures occurred at different time can last different duration. For example, under strong and sustained hurricane wind, failures that do not happen in day-to-day operation can occur due to, e.g., falling debris and power lines. We shall further elaborate this through the real-life example in Section \ref{sec:RealData}.

A failure process and a recovery process are related by failure durations. As a special case when failure follows a non-homogeneous Poisson process, the following theorem shows that the recovery process is also a non-homogeneous Poisson process parameterized by recovery rate $\lambda_r(t)$. In addition $\lambda_r(t)$ can be expressed by failure rate $\lambda_f(t)$ with the help of the distribution of failure duration $g(d|t)$.

\indent

\textbf{Theorem}\label{thm:rp}
\textit{Assume \\
(a) Independent failure occurrence $\{T\}$ obeys a non-homogeneous Poisson process $\{N_f(t)\}$ with intensity function $\lambda_f(t)$; \\
(b) Failure duration $\{D\}$ follows a conditional probability density function $g(d|t)$ for $d \ge 0$, $t \ge 0$. \\
Then the recovery time $\{T+D\}$ is drawn from a non-homogeneous Poisson process $\{N_r(t)\}$ with recovery intensity function,
\begin{equation}
\lambda_r(t)=\int_0^{t} \lambda_f(s)g(t-s|s)ds.
\label{eq:RRate}
\end{equation}
}
The proof of the theorem is given in Appendix \ref{app:Proof}.

Hence, if a failure process is non-homogeneous Poisson, our general formulation of the temporal random process reduces to a $Mt/Gt/\infty$ queue. $Mt$ indicates time-dependent Poisson failures. $Gt$ indicates non-stationary distribution of failure duration. $\infty$ is for the infinite number of servers for repair; thus recovery can occur right after failure. Furthermore, when $g(d|t)=g(d)$ is stationary, $\lambda_r(t)$ reduces to the convolution of the failure rate and service time distribution $g(d)$. The $Mt/Gt/\infty$ model reduces to $Mt/G/\infty$ queue developed in \cite{Massey93}.

\subsection{Special Cases}

We consider two special cases where failure, recovery and joint processes exhibit simple analytical expressions. These expressions provide insights to an entire non-stationary life cycle of failure-and-recovery process under different failure scenarios.

\indent

\textit{Case 1: Failure and recovery in day-to-day operation}:

When there are no significant external disruptions, power outages are assumed to occur randomly and sporadically. The failure intensity remains at a constant level, i.e., $\lambda_f(t)=\lambda_0$, where $\lambda_0 \ge 0$ does not vary with time. The number of failures $N_f(t)$ thus follows a homogeneous Poisson process in day-to-day operation. The recovery rate in day-to-day operation $\lambda_{r,0}(t)$ in Eq.\ref{eq:RRate} reduces to,
\begin{equation}
\lambda_{r,0}(t)=\lambda_0 \int_{0}^{t} g(t-s|s)ds.
\end{equation}
As $t\rightarrow \infty$, $\lambda_{r,0}(t)\rightarrow \lambda_0$. This suggests that the recovery process in day-to-day operation is also a homogeneous Poisson process at a long time horizon. Then the expectation $E\{N_0(t)\}$ of joint failure-and-recovery process in day-to-day operation is
\begin{equation}
E\{N_0(t)\}=\int_0^t [\lambda_0 - \lambda_{r,0}(v) ]dv.
\end{equation}

\indent

\textit{Case 2: Surging failure during a natural disaster}:

Now consider a disaster scenario where failures occur suddenly and intensely. $\lambda_f(t)$ increases from a small value $\lambda_0$ to a large value in a short time duration $0 \le t < t_1$. The failure rate can then be written as,
\begin{equation}
\lambda_f(t)=\lambda_m(t)[u(t)-u(t-t_1)]+\lambda_0,
\label{eq:FRateSurg}
\end{equation}
where $u(t)$ is the unit step function, $\max_{t} \lambda_m(t) \gg \lambda_0$. This corresponds to an extreme case when a disaster causes sudden failures at time $0$ and then weakens right after.

$\lambda_r(t)$ in Equation \ref{eq:RRate} becomes,
\begin{equation}
\lambda_r(t) = \int_{0}^{\min(t,t_1)} \lambda_m(s)g(t-s|s)ds+\lambda_{r,0}(t).
\label{eq:RRateSurg}
\end{equation}

When $t \gg t_1$, $\lambda_r(t) \approx \lambda_0$. Furthermore, when $t_1$ is sufficiently small, i.e., a surge of failures upon the disaster only lasts a short time,
\begin{equation}
\lambda_r(t) \doteq \lambda_m(0) g(t|0) \min\{t,t_1\} u(t) + \lambda_{r,0}(t).
\label{eq:RRateSurg}
\end{equation}
When the peak of surging failures $\lambda_m(0) \gg \lambda_0$, the recovery rate in Eq.\ref{eq:RRateSurg} is approximately proportional to the surging failure rate and distribution of recovery time immediately after the disaster. In the long run, $\lambda_r(t)$ reduces to $\lambda_0$.

Substituting Eq.\ref{eq:FRateSurg} and \ref{eq:RRateSurg} to Eq.\ref{eq:frp}, we have the expectation of the failure-and-recovery process in surging failures as,
\begin{equation}
E\{N(t)\} \doteq \lambda_m(0)[1-G(t|0)]\min\{t,t_1\}+ E\{N_0(t)\},
\label{eq:frpSurg}
\end{equation}
where $G(t|s)=\int_{0}^t g(v|s) dv$ is the conditional cumulative density function (cdf) of failure duration given time $t$.

\indent

Additional characteristics of recovery can further reduce the above expressions. When infant recovery completely dominates a recovery process, $g(d|0)=0$ when $d>d_0$. Hence, for an impulse-like surge of failures at $t=0$, the resulting recovery from the surge lasts $d_0$ duration, i.e.,
\begin{equation}
\medmath{\lambda_r(t) \doteq \lambda_m(0) g(t|0) \min\{t,t_1\} [u(t)-u(t-d_0)] + \lambda_{r,0}(t)}.
\label{eq:RRateSfIr}
\end{equation}
Eq.\ref{eq:frpSurg} in dominating infant recovery becomes,
\begin{equation}
\medmath{
\begin{split}
&E \{N(t)\} \doteq \\
&
\left\{
\begin{array}{ll}
  \lambda_m(0) [1-G(t|0)] \min\{t,t_1\} + E\{N_0(t)\},  &  0 \le t < d_0, \\
  E\{N_0(t)\},  &  t \ge d_0. \\
\end{array}
\right.
\end{split}
}
\label{eq:frpSfIr}
\end{equation}

Here failure duration $d_0$ is assumed to be larger than the end of failure process $t_1$ to simplify the expression. Eq.\ref{eq:RRateSfIr} and Eq.\ref{eq:frpSfIr} show that when a recovery process consists of only infant recovery, all failures due to disasters recover by $d_0$ after the failure eruption. After $d_0$ the distribution network resumes day-to-day operation.

When aging recovery dominates the recovery process, $g(d|0) \approx 0$ for $d < d_0$. This implies that recovery begins with delay $d_0$ after a surge of failures. The corresponding recovery rate reduces from Equation \ref{eq:RRateSurg}, where
\begin{equation}
\lambda_r(t) \doteq \lambda_m(0) g(t|0) \min\{t,t_1\} u(t-d_0) + \lambda_{r,0}(t).
\label{eq:RRateSfAr}
\end{equation}
Eq.\ref{eq:frpSurg} in dominating aging recovery becomes,
\begin{equation}
\begin{split}
&E \{N(t)\} \doteq \\
&\left\{
\begin{array}{ll}
    \lambda_m(0) \min\{t,t_1\} + E\{N_0(t)\}, & 0 \le t < d_0, \\
    \lambda_m(0) t_1 [1-G(t|0)] + E\{N_0(t)\}, & t \ge d_0. \\
\end{array}
\right.
\end{split}
\label{eq:frpSfAr}
\end{equation}
Here we also assume $d_0 > t_1$. Eq.\ref{eq:frpSfAr} shows another extreme case where recovery does not begin until $d_0$ delay from the failure eruption. Then the failures start to recover slowly. Hence, aging recovery shows the difficulty in resuming service to customers, and is thus an undesirable characteristic for the smart grid.

In the general failure-and-recovery process, recovery begins as soon as failures occur. During the disaster, the failure process dominates and $E\{N(t)\}$ increases rapidly. Afterwards, the recovery process dominates. At the large time scale and when disaster lasts for a short period of time, the overlap between the failure process and recovery process can be neglected. Thus the failure-and-recovery process splits into individual processes, failure and recovery, respectively.


\section{Resilience}\label{sec:Resilience}

We now derive network resilience using the pertinent parameters for an entire life cycle of non-stationary failure and recovery.

\subsection{Definition}

An intuition on how to define resilience results from the previous section. Resilience should be characterized by virulence of failures $\lambda_f(t)$, speed of recovery $g(d|t)$ and threshold $d_0$. These three parameters determine infant recovery. A distribution network experiences a combination of infant and aging recovery in general. A network is intuitively more resilient when exhibiting more infant recovery. Hence we define the resilience as the probability of infant recovery.

\indent

\textbf{Definition}: Resilience: Given a threshold value $d_0$ on failure duration, the resilience of a power distribution network is defined as \begin{equation}
s(d_0)=P\{D<d_0\}.
\label{eq:SDef}
\end{equation}

\indent

At a high-level, $s(d_0)$ measures the resilience of the grid from large-scale disruptions, and reflects the ability for a distribution network as a whole to survive large-scale failures. Here we use $P\{D<d_0\}$ rather than time dependent conditional probability $P\{D<d_0|t\}$. This is because resilience, when viewed as a network-wide quantity, should include an entire life-cycle of failure and recovery but not depend on specific time epoch $t$.

\subsection{Resilience Parameters}

Resilience can be expressed explicitly by the three pertinent parameters as follows,
\begin{equation}
\begin{split}
s(d_0)=&E_{t} \{G(d_0|t)\} \\
      =&\int_t P\{D<d_0|t\}f(t) dt \\
      =&\int_{t=0}^{\infty} \int_{v=0}^{d_0}g(v|t)f(t) dv dt,
\label{eq:SAltDef}
\end{split}
\end{equation}
where $ G(d_0|t)= P\{D<d_0|t\}$, and $f(t)$ is the probability density function of failure time given in Equation \ref{eq:pdfFP}. Eq.\ref{eq:SAltDef} results in an alternative expression for resilience. There, $s(d_0)$ is the expected value of the probability of infant recovery $G(d_0|t)$ averaged over the non-stationary failure process. This expression shows explicitly how resilience is determined by the three pertinent parameters.


\subsection{Threshold}

$d_0$ is one other pertinent parameter that determines the resilience. For a given value of $d_0$, if infant recovery dominates the recovery process over the entire failure duration, $s(d_0) \rightarrow 1$. If aging recovery dominates the recovery process over the entire failure process, $s(d_0) \rightarrow 0$. Thus, a larger $s(d_0|t)$ represents a more resilient power distribution network. When $s(d_0)=0.5$, the infant and the aging recovery take equal weight.

How to determine the value for $d_0$? $d_0$ can result from practical considerations or customer requirements. For example, when recovering within $24$ hours is regarded as acceptable for a disaster scenario, $d_0=24$ defines infant recovery.

\begin{figure}
\begin{center}
\includegraphics[width=0.5\textwidth]{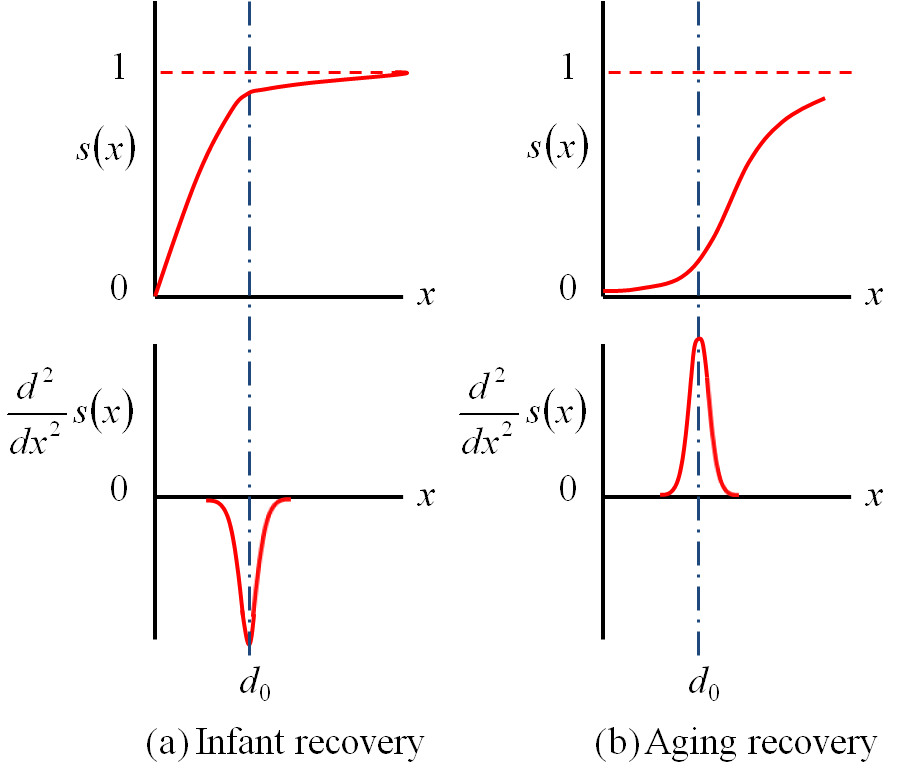}
\caption{Illustration: quantitative method for determining $d_0$.}
\label{fig:D0}
\end{center}
\end{figure}

A value for $d_0$ can also be determined more objectively. An intuition results from the fact that if a network is dominated by infant recovery, most failure durations are less than $d_0$. Hence, the slope of $s(x)$ is relatively large for $x<d_0$ and relatively small for $x>d_0$. This implies that $d_0$ corresponds to a pertinent change point of the slope of $s(x)$, i.e., a deep valley in the second derivative $\frac{d^2}{dx^2}s(x)$. This is illustrated in in Fig.\ref{fig:D0}(a). In contrast, when a network is dominated by aging recovery, most of the failure durations are larger than $d_0$. Thus the slope of $s(x)$ is small for $x< d_0$ and large for $x>d_0$. $d_0$ then corresponds to a positive peak in $\frac{d^2}{dx^2}s(x)$ as illustrated in Fig.\ref{fig:D0}(b). Hence, $d_0$ is determined by the largest magnitude of the second derivative of the $s(x)$,

\begin{equation}
\begin{split}
d_0 =
\left\{
    \begin{array}{ll}
    \text{argmin}_{x} \{\frac{d^2}{dx^2}s(x)\}, & \text{$s(x)$ is concave},  \\
    \text{argmax}_{x} \{\frac{d^2}{dx^2}s(x)\}, & \text{$s(x)$ is convex}.  \\
    \end{array}
\right.
\end{split}
\label{eq:d0}
\end{equation}
We shall provide an example using real data in Section \ref{sec:RealData}.

\section{Large-Scale Outages due to Hurricane Ike}\label{sec:RealData}

We now apply the above framework on non-stationary failure-recovery processes to a real-life example of large-scale utility-service disruptions caused by a hurricane. Our focus is on using real data to learn the three resilience parameters $\lambda_f(t)$, $g(d|t)$ and $d_0$ and then to estimate the resilience of an operational power distribution network.

\subsection{Real Data and Processing}

Hurricane Ike was one of the strongest hurricanes occurred in 2008. Ike caused large scale power failures, resulting in more than 2 million customers without electricity, and marked as the second costliest Atlantic hurricane of all time \cite{Blake11}\cite{fema08}.

Reported by National Hurricane Center \cite{nhc_ike_report}, the storm started to cause power outages across the onshore areas in Louisiana and Texas on September 12, 2008 prior to the landfall. Ike then made a landfall at Galveston, Texas on 2:10 a.m. Central Daylight Time (CDT), September 13, 2008, causing strong winds, flooding, and heavy rains across Texas. The hurricane weakened to a tropical storm at 1:00 p.m. September 13 and passed Texas by 2:00 a.m. September 14.

\begin{figure}
\begin{center}
\includegraphics[width=0.5\textwidth]{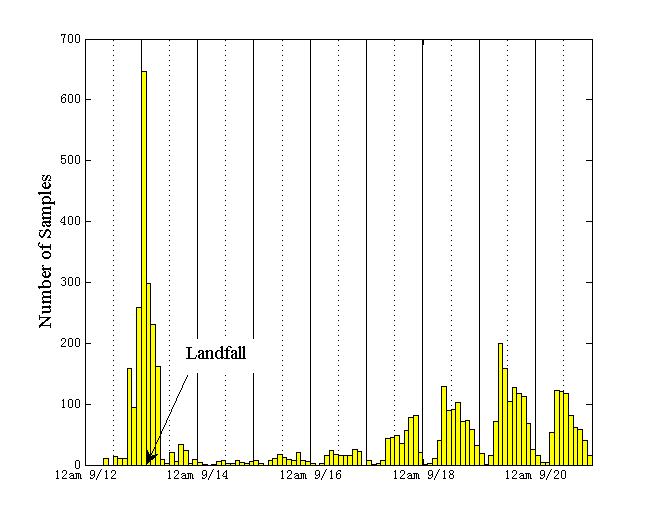}
\caption{Histogram of failure occurrence times during Hurricane Ike.}
\label{fig:HistPowerIniT1}
\end{center}
\end{figure}

Widespread power outages were reported across Louisiana and Texas starting September 12 \cite{fema08}. A major utility provider collected data on power outages from more than ten counties. The outages include various component failures in a distribution network such as failed circuits, fallen poles, and non-operational substations. The raw data set consists of 5152 samples. Each sample consists of the failure occurrence time ($t_i$) and duration ($d_i$) of a component ($i$) in a distribution network. The accuracy for time $t$ is a minute. The data set contains failures occurred from September $12$ through $14$, 2008.

The 5152 samples on the failure occurrence time are plotted in Fig.\ref{fig:HistPowerIniT1}, where the length of each bin is two hours. There were significantly more failures occurred from 7 a.m. September 12 to 4 a.m. September 14, during which Hurricane Ike made the landfall in Texas. There are 2005 samples in this time period. Those 2005 samples are considered as failures due to Hurricane Ike.

Furthermore, the data set contains groups of failures that occurred within a minute. As a minute is the smallest time scale for each sample, the groups are considered as dependent failures. Dependent failures are grouped as one failed entity ($i$), with a unique failure occurrence time $t_i$ and duration $d_i$. After such preprocessing, the resulting data set has 465 failed entities. Two outliers with negative failure duration are further removed. The remaining 463 failed entities from 7 am September 12 to 4 am September 14 are referred to as nodes. $D=\{t_i, d_i\}_{i=1}^{463}$ is the data set we use for the rest of the paper.

\subsection{Empirical Failure Process}

We now focus on studying the empirical failure process using data set.

\subsubsection{Learning failure rate}

\begin{figure}
\begin{center}
\includegraphics[width=0.5\textwidth]{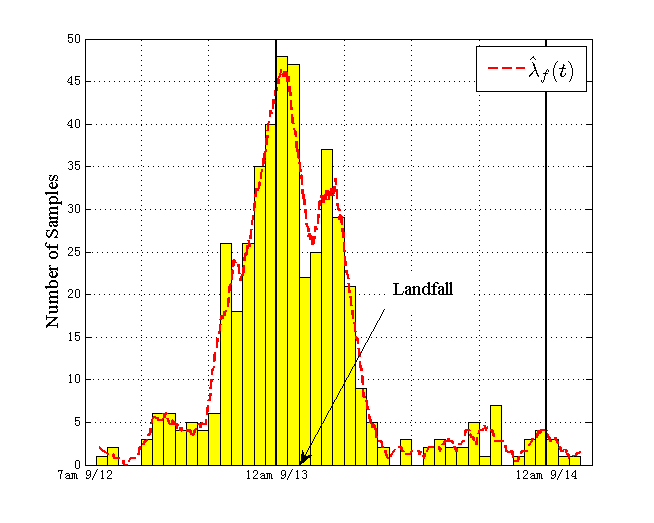}
\caption{Histogram of failure occurrence times and the failure rate $\lambda_f(t)$ during Hurricane Ike.}
\label{fig:HistPowerIniT2}
\end{center}
\end{figure}

First, we use the data set to learn failure rate function $\lambda_f(t)$. The empirical rate function is estimated using a simple moving average \cite{Trees71}: $\hat{\lambda}_f(t)=\frac{\hat N_f(t+\tau)-\hat N_f(t-\tau)}{2\tau}$, where $\tau>0$. The resulting rate function is overlaid with the samples on the number of failures $\hat N_f(t)$ in Figure \ref{fig:HistPowerIniT2}, where each bin is of duration 1 hour, $\tau=5$ hours.

The failure rate function shows a time-varying, i.e., non-stationary, intensity of new failure occurrence, and can be described as follows.
\begin{itemize}
\item Prior to 7 p.m. September 12, the intensity was low, i.e., fewer than 5 new failures occurred per hour. Hence $\lambda_0=5$ where $\lambda_0$ is considered as the failure rate in day-to-day operation.
\item At 7 p.m. September 12, the intensity increased sharply first to 25 new failures per hour. In the next 6 hours, the intensity reached the peak value of nearly 50 new occurrences per hour. This is consistent to the weather report \cite{nhc_ike_report} that the strong wind about 145 mph and flooding impacted the onshore areas even prior to the landfall. The time when the peak occurred coincides with the landfall at 2:10 a.m 9/13 CDT.
\item After staying at the high level for about 12 hours (from 7 p.m. September 12 to 7 a.m. September 13), the intensity decreased rapidly back to a low level of less than 5 new failures per hours.
\end{itemize}


\subsubsection{Non-Homogeneous Poisson Model}

We now consider a hypothesis $H_0$ that these $463$ failure occurrences are governed by a non-homogeneous Poisson process (NHPP). We perform Pearson's test \cite{Plackett83} on hypothesis $H_0$ that the empirical failure process $\hat N_f(t)$ follows a non-homogeneous Poisson Process (NHPP) with intensity function $\hat{\lambda}_f(t)$. The test is on two aspects: (a) independence of the outages occurred at the large time scale of minutes, and (b) the sample mean, i.e., the empirical rate function $\hat{\lambda}_f(t)$, is sufficient for characterizing the failure process.

We divide the time duration from 7 a.m. September 12 to 4 p.m. September 14 into 400 intervals. In each interval, the number of new failures is compared with the expectation. The sum of the square errors from all intervals results in a chi-square statistic. The chi-square statistic is $\chi^2=0.79$, with a degree of freedom of 2. Given a confidence level at $95\%$, a threshold value is obtained as $\chi_{0.05,2}^2 = 5.99$, where $Pr(\chi^2<\chi_{0.05,2}^2)=0.95$. The chi-square statistic $\chi^2$ obtained from the data is below the threshold $\chi^2<\chi_{0.05,2}^2$. Hence hypothesis $H_0$ is not rejected. The detailed procedure of Pearson's test is given in Appendix \ref{app:Htest}.

\begin{figure}
\begin{center}
\includegraphics[width=0.5\textwidth]{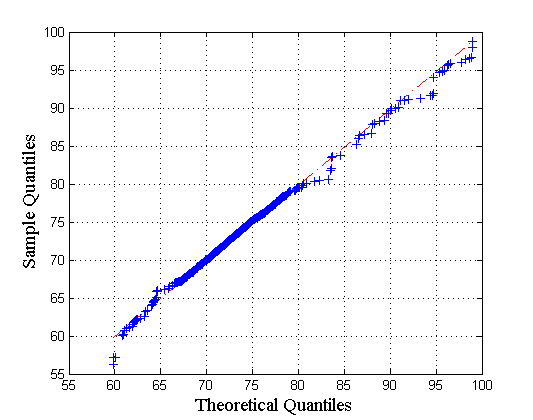}
\caption{Quantile-Quantile plot on occurrence time of power outages.}
\label{fig:QQplotPIniT}
\end{center}
\end{figure}

However, not rejecting $H_0$ is insufficient for accepting the hypothesis. The goodness of fit of NHPP to the data is further validated through Quantile-Quantile (QQ) plot given in Figure \ref{fig:QQplotPIniT}. There, the samples are compared with an non-homogeneous Poisson process with intensity function $\hat \lambda_f(t)$. The figure shows that the non-homogeneous Poisson process with the learned intensity function indeed exhibits a good fit to the data. Based on the result from Pearson's hypothesis testing and the QQ plot, these 463 power failures occur independently, and obey a non-homogeneous Poisson distribution.

\subsection{Empirical Recovery Process}

We now learn the empirical recovery process characterized by $g(d|t)$, the conditional probability density function of failure duration given failure occurrence time. Our objective here is to identify infant and aging recovery.


\subsubsection{Data}

The $463$ samples on the failure durations in our data set correspond to the failures occurred from 7 a.m. September 12 to 4 p.m. September 14. These samples result in a joint empirical distribution $\hat g(d,t)$ depicted in the 3-D Fig.\ref{fig:HistPDur}. Each bin is of length (initial failure time) of $1$ hour and width (duration) of $4$ hours. The height of each bin located at initial failure time $t$ and failure duration $d$, represents the number of failures that occur at $t$ and last for $d$.



\indent

\subsubsection{Mixture Model}

Given failure time $t$, we select a mixture model as the probability density function $g(d|t)$ for duration $d >0$,
\begin{equation}
g(d|t)=\sum_{j=1}^{l(t)}{\rho_j (t)g_{j}(d|t)},
\end{equation}
where $l(t)$ is the number of mixtures at time $t$, $\rho_j(t)$ ($1 \le j \le l$) is a weighting factor for the $j$th mixture function $ g_j(d|t)$, and $\sum \rho_j(t)=1$. All these quantities vary with failure time $t$ for a non-stationary recovery process.

A mixture model is chosen since its parameters exhibit interpretable physical meaning. A parametric family of Weibull mixtures is particularly appealing as the parameters correspond to infant and aging recovery directly \cite{Ross10}. Weibull distributions have been widely used in survival analysis \cite{Kalbfleisch02}\cite{Hosmer08} and reliability theory \cite{Ross10}, but not in characterizing recovery from large-scale disturbances and resilience parameters. Specifically, a Weibull distribution is
\begin{equation}
\medmath{
g_j(d|t;\gamma_j(t),k_j(t))=\frac{k_j(t)}{\gamma_j(t)} (\frac{d}{\gamma_j(t)})^{k_j(t)-1} e^{-(\frac{d}{\gamma_j(t)})^{k_j(t)}},
}
\end{equation}
where $d>0$, $k_j(t)$ and $\gamma_j(t)$ are the shape and scale parameters respectively. A shape parameter $k_j(t)$ is especially important for determining the type of recovery signified by a mixture component. The smaller $k_j(t)$ is, the faster the decay rate of $g_{j}(d|t)$, the shorter the failure duration and thus the faster the recovery. Hence, $ k_j(t) < 1$ corresponds to infant recovery whereas $ k_j(t) > 1$ corresponds to aging recovery. Weighting factor $\rho_j (t)$ signifies the importance of a component $g_{j}(d|t)$. For a non-stationary recovery process, these parameters are varying with failure time $t$.

\indent

\subsubsection{Learning Mixture Parameters}

The parameters of the mixtures of Weibull distribution are learned from the data. For simplicity, we use a piecewise homogeneous function to approximate $g(d|t)$. We divide the failure time into $m$ intervals. Within an interval $\psi_i$, $g(d|t\in \psi_i)=g_i(d)$ is assumed to be a time homogeneous function that does not vary with failure time $t$. For different intervals, $g(d|t \in \psi_i)$'s have different parameters for non-stationarity, where
\begin{equation}
g(d|t \in \psi_i)=\sum_{j=1}^{l_i} \rho_{i,j} g_{i,j}(d;\gamma_{i,j},k_{i,j}).
\label{eq:wblpdf}
\end{equation}

\begin{table}
  \centering
  \caption{Estimated parameters of the five homogeneous Weibull distributions.} \label{tab:wblpar}
  \begin{tabular}{c|c|c|c|c|c}
      \hline
      $g_1(d)$ & 1 & 2 & 3 &  & $P\{d<13|\psi_1\}$ \\
      \hline
      $\rho_{1,j}$    & 0.486  & 0.257  & 0.257 &  & \\
      $\gamma_{1,j}$  & 0.710  & 14.400 & 211.830 &  & $55.99\%$ \\
      $k_{1,j}$       & 1.000   & 10.533 & 10.679 &  & \\
      \hline
      $g_2(d)$ & 1 & 2 & 3 & 4 & $P\{d<13|\psi_2\}$ \\
      \hline
      $\rho_{2,j}$    & 0.321  & 0.206  & 0.019 & 0.454  & \\
      $\gamma_{2,j}$  & 2.680  & 7.640 & 21.220 & 173.580  & $47.16\%$ \\
      $k_{2,j}$       & 0.370  & 2.910 & 46.230 & 3.090  & \\
      \hline
      $g_3(d)$ & 1 & 2 & 3 &  & $P\{d<13|\psi_3\}$ \\
      \hline
      $\rho_{3,j}$    & 0.143  & 0.472  & 0.385 &  & \\
      $\gamma_{3,j}$  & 0.530  & 12.300 & 135.072 &  & $56.89\%$ \\
      $k_{3,j}$       & 2.500  & 15.201 & 4.424 &  & \\
      \hline
      $g_4(d)$ & 1 & 2 &  &  & $P\{d<13|\psi_4\}$ \\
      \hline
      $\rho_{4,j}$    & 0.323  & 0.677  & &  & \\
      $\gamma_{4,j}$  & 11.041 & 112.245 & &  & $29.27\%$ \\
      $k_{4,j}$       & 5.310  & 12.398 & &  & \\
      \hline
      $g_5(d)$ & 1 & 2 & 3 &  & $P\{d<13|\psi_5\}$ \\
      \hline
      $\rho_{5,j}$    & 0.273 & 0.159  & 0.568 &  & \\
      $\gamma_{5,j}$  & 2.479 & 21.555 & 134.053 &  & $32..60\%$ \\
      $k_{5,j}$       & 0.987 & 1.702  & 5.070 &  & \\
      \hline
  \end{tabular}
\end{table}

Referring to Fig.\ref{fig:HistPDur}, we divide the time into 5 intervals, $\{\psi_i, i=1,2,3,4,5\}$, the boundaries of each interval are depicted by dashed lines in Fig.\ref{fig:HistPDur}.

\begin{figure}
\begin{center}
\includegraphics[width=0.5\textwidth]{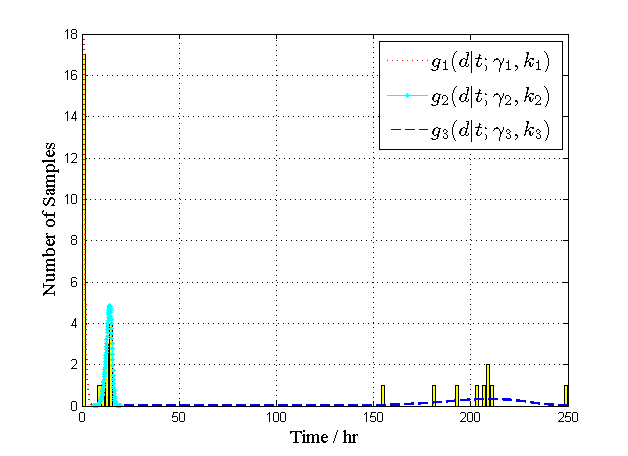}
\caption{Empirical distribution of failure duration corresponding to the failures occurred in $\psi_1$.}
\label{fig:PDWbl}
\end{center}
\end{figure}

Within each interval, the parameters of the mixture Weibull distribution are learned through maximum likelihood estimation \cite{LeCam90}, and shown in Table \ref{tab:wblpar}. Each mixture represents one cluster of failure durations. For example, distribution of durations in $\psi_1$ is depicted in Fig.\ref{fig:PDWbl}, where we observe 3 clusters.

For different intervals, the failure duration exhibits distinct distribution. For example, $\psi_1$ (7 a.m. September 12 to 7 p.m. September 12) corresponds to the time before hurricane. In this interval, the network was not yet impacted by hurricane Ike and was under day-to-day operation. Most failure durations were as short as a few hours. $\psi_2$ (7 p.m. September 12 to 3 a.m. September 13) corresponds to the time right before the landfall and hurricane began to cause large-scale failures. In this interval, more prolonged failures occur and recovery became more difficult than day-to-day operation.

\subsection{Overall Failure-and-Recovery Process}

\begin{figure}
\begin{center}
\includegraphics[width=0.5\textwidth]{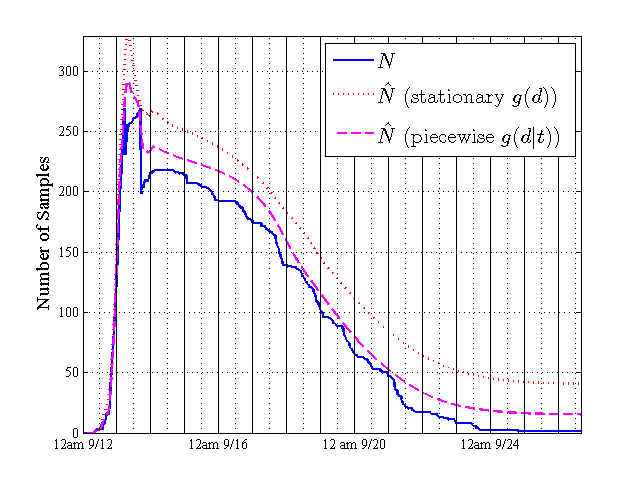}
\caption{Comparison between the joint failure-recovery process $N(t)$ from the data set and the reconstructed process $\hat N(t)$ using learned parameters.}
\label{fig:FRP}
\end{center}
\end{figure}

We now compare the empirical failure-and-recovery process $\{N(t)\}$ with the learned process $\hat N(t)$. $N(t)$ is obtained by directly adding up the number of failed nodes from the actual data. $\hat N(t)$ is obtained by reconstructing the temporal process with learned $\hat \lambda_f(t)$ and $\hat \lambda_r(t)$ through Eq.\ref{eq:frp}. Fig.\ref{fig:FRP} shows the comparisons. The dotted $\hat N(t)$ is obtained with the assumption that $g(d)$ is stationary over time. The dashed $\hat N(t)$ is obtained with the piecewise stationary $g(d|t)$ given in Table \ref{tab:wblpar}. All estimated processes are able to capture the trend in the data. However, the stationary distribution of failure durations $g(d)$ deviates significantly from the actual sample path $N(t)$. This shows that the piecewise stationary $g(d|t)$ better approximates the actual failure-and-recovery process.

\subsection{Resilience}

\begin{figure}
\begin{center}
\includegraphics[width=0.5\textwidth]{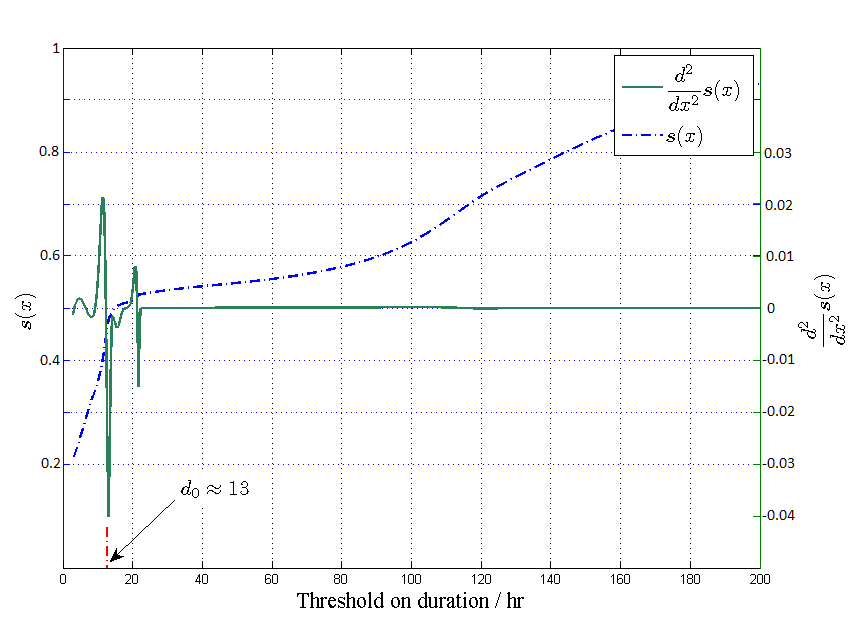}
\caption{Resilience curve $s(x)$ of the power network.}
\label{fig:s}
\end{center}
\end{figure}

To evaluate resilience, we first calculate the probability of infant recovery within each interval $\psi_i$ as shown in Table \ref{tab:wblpar}. The resilience curve $s(x)$ is then obtained by averaging these probabilities over failure time. Fig.\ref{fig:s} depicts $s(x)$ together with its second derivative $\frac{d^2}{dx^2}s(x)$. $s(x)$ is a concave function. Hence, the threshold is obtained as,


\begin{equation}
d_0= \text{argmin}_{x}\{\frac{d^2}{dx^2}s(x)\}\approx 13.
\end{equation}
we compute the resilience
\begin{equation}
s(13)=E_{t}\{G(13|t)\}=0.4722.
\label{eq:s13}
\end{equation}

Hence, $47.22\%$ of the recoveries occurred within 13 hours whereas $52.78\%$ occurred later than 13 hours. Such resilience close to 0.5 reveals the combined nature of infant and aging recovery in the distribution network. Specifically, $s(13)<0.5$ indicates a slight dominance of aging recovery over infant recovery.


\section{Discussion and Conclusion}\label{sec:Conclusion}
A non-stationary random process has been derived to model large-scale failure and recovery of a power distribution network under an external disturbance. The non-stationary network failure and recovery are characterized by time-varying failure rate and probability distribution of recovery time. These two quantities, together with a threshold on recovery time, define network resilience as the probability of rapid recovery. Analytical expressions have been derived to characterize the non-stationary failure and recovery under extreme conditions.

Real data has been obtained from an operational distribution network for a large-scale external disturbance, Hurricane Ike. Model parameters of the failure-recovery process have been estimated through non-stationary learning using the real data. Pearson's statistical test is applied to validate the assumptions of the failure model. The resilience metric has been evaluated, and shown for the operational network that $47\%$ of $5000+$ failures recovered within $13$ hours whereas the remaining prolonged as long as 12 days. This provides a quantitative measure of the resilience of the distribution network and a baseline for improvement.

How to improve resilience to reduce failures needs to be further incorporated in the resilience metric. Non-stationary learning algorithms need to be further developed for on-line learning of time-varying data.

\section{Acknowledgement}

The authors would like to thank Chris Kung, Jae Won Choi, Daniel Burnham and Xinyu Dai at Georgia Tech for data processing, Anthony Kuh at University of Hawaii for helpful discussions on distribution networks. Support from National Science Foundation (ECCS 0952785) is gratefully acknowledged.

\appendices
\section{}\label{app:Proof}

\textbf{Proof of Theorem:}
In order to prove that the recovery process $\{N_r(t)\}$ is a Poisson process, we just need to show the increment of recovery process, i.e., recoveries occurs in $(t, t+\tau)$, is an independent Poisson random variable. Now consider non-overlapping intervals $O_1$, ..., $O_k$. We say a failure is type $i$, $i=1,...,k$, if it recovers in the interval $O_i$. The number of the type $i$ events equals to the number of recoveries occur in the interval $O_i$. By Proposition 5.3 in \cite{Ross10}, it follows that the number of recoveries occur in the interval $O_i$, i.e., the increment of recovery process, is an independent Poisson random variable with mean,
\begin{equation}
E\{\text{recoveries in $O_i$}\} = \int_0^t \lambda_f(s) P_i(s) ds,
\label{eq:proof1}
\end{equation}
where $P_i(s)$ is the probability that a failure is type $i$. Thus, we prove that $\{N_r(t)\}$ is a Poisson process.

Then we compute $\lambda_r(t)$ and show the non-homogeneity. Consider interval $(t, t+\tau)$ and a failure occurs at time $s$, where $s<t$. The probability that this failure recovers during $(t, t+\tau)$ is $G(t+\tau-s|s)-G(t-s|s)$. Eq.\ref{eq:proof1} becomes,
\begin{equation}
\begin{split}
&E\{N_r(t+\tau)-N_r(t)\} \\
=&\int_0^t \lambda_f(s)[G(t+\tau-s|s)-G(t-s|s)]ds \\
=&\tau \int_0^t \lambda_f(s)g(t-s|s) ds + o(\tau), \\
\end{split}
\label{eq:proof2}
\end{equation}
where $G(t+\tau-s|s)-G(t-s|s)=g(t-s|s)\tau + o(\tau)$. We can rewrite Eq.\ref{eq:proof2} as,
\begin{equation}
\int_0^t\lambda_f(s)g(t-s|s)ds =
\lim_{\tau \rightarrow 0} \frac{E\{N_r(t+\tau)-N_r(t)\}}{\tau}.
\end{equation}
The right-hand-side is just $\lambda_r(t)$ (Eq.\ref{eq:RRateDef}). Hence, the recovery rate function $\lambda_r(t)$ is,
\begin{equation}
\lambda_r(t)=\int_0^t \lambda_f(s)g(t-s|s)ds.
\end{equation}
Due to the non-homogeneity of $\{N_f(t)\}$, $\lambda_f(t)$ is a time-varying function, so $\lambda_r(t)$ is also time-varying function. Hence, we show the non-homogeneity of $\{N_r(t)\}$, which completes the proof. $\blacksquare$

\section{}\label{app:Htest}

\textbf{Pearson's Hypothesis Test:} 
The hypothesis test is based on a chi-square statistic which compares the failure occurrence times with their sample mean. The details of testing $H_0$ that failure occurrence times are drawn from a NHPP $\{N_f(t)\}$ are given here.

\textbf{1.} Compute the estimated failure rate $\hat \lambda_f(t)$. At time $t$, count the number of failure occurrences in $(t-\tau,t+\tau)$, then $\hat \lambda_f(t)=\frac{N_f(t+\tau)-N_f(t-\tau)}{2\tau}$.

\textbf{2.} Divide the failure occurrence times into $m$ intervals. Count the number of failure occurrences in each interval. Let $c_i$ denote the number of occurrence in interval $i$, where $i=1,2,...,m$. Let $k=\max{c_i,i=1,2,...m}$.

\textbf{3.} Count $O_j$, which is the observed number of intervals with $j$ failure occurrences, where $j=0,1,...,k$.

\textbf{4.} Use the estimated $\hat \lambda_f(t)$ to compute $E_j$, which is the expected number of intervals with $j$ failure occurrences. $E_j=\sum_{i=1}^{m}\frac{e^{\hat \lambda_{s_i,t_i}} \hat \lambda_{s_i,t_i}^{j}}{j!}$, where $(s_i,t_i)$ is the $i$th time interval, $\hat \lambda_{s_i,t_i}=\int_{s_i}^{t_i} \hat \lambda_f(t)dt$.

\textbf{5.} Compute the sum $\chi^2=\sum_{j=0}^k{\frac{(O_j-E_j)^2}{E_j}}$. $\chi^2$ is a chi-square statistic, with degree of freedom $dof=k-$(number of independent parameter fitted)$-1$. Since one parameter $\hat \lambda_f(t)$ is fitted, the degree of freedom is $k-2$.

\textbf{6.} Given a confidence level, for instance $95\%$, we obtain a threshold value $\chi_{0.05,dof}^2$. The hypothesis $H_0$ is rejected if $\chi^2<\chi_{0.05,dof}^2$; otherwise, $H_0$ cannot be rejected.


\bibliographystyle{IEEEtran}
\bibliography{yunref}

\end{document}